\documentstyle[11pt,ask,epsf]{article}
\makeatletter
\@addtoreset{equation}{section}
\makeatother

\def\mr{\rm}	% {\mathrm}
\newcommand{\cA}{{\cal A}}
\newcommand{\cD}{{\cal D}}

\newcommand{\cL}{{\cal L}}

\newcommand{\cU}{{\cal U}}
\newcommand{\cZ}{{\cal Z}}

\newcommand{\half}{\mbox{\small $\frac{1}{2}$}}
\newcommand{\ihalf}{\mbox{\small $\frac{i}{2}$}}
\def\imag{\mathop{\mr Im}}

\newcommand{\quart}{\mbox{\small $\frac{1}{4}$}}

\def\real{\mathop{\mr Re}}

\newcommand{\sixth}{\mbox{\small $\frac{1}{6}$}}
\newcommand{\third}{\mbox{\small $\frac{1}{3}$}}

\def\Det{\mathop{\mr Det}}
\def\tr{\mathop{\mr tr}}
\def\Tr{\mathop{\mr Tr}}
\newcommand{\vdot}{\!\cdot\!}
\def\zahlen{{\sf Z\kern-0.4em Z}}

\def\dfrac#1#2{{\displaystyle \frac{#1}{#2}}}
\def\tfrac#1#2{{\textstyle \frac{#1}{#2}}}
\preprint{FERMILAB-PUB-95/073-T \\
{\tt hep-lat/9504007} % \\
}
\title{Chiral Fermions Coupled to Lattice Gauge Fields}
\author{Andreas S. Kronfeld \\[1.5em]
Theoretical Physics Group, Fermi National Accelerator Laboratory, \\
P.O. Box 500, Batavia, IL 60510}
\date{15 April 1995}
\begin{document}
\setcounter{page}{0}
\maketitle
%-----------------------------------------------------------------------
\vfill
\begin{abstract} \normalsize
Lattice fermions have well-known difficulties with chiral symmetry.
To evade them it is possible to couple {\em continuum\/} fermions
to {\em lattice\/} gauge fields, by introducing an interpolation of the
latter.
Following this line of thinking, this paper presents two Euclidean
formulations of the effective action that appears after functional
integration over fermion fields, one for vector-like and the other
for chiral couplings.
With suitable finite-mode regulators both effective actions can be
evaluated in a finite computation.
The prescriptions provided here contain some details not found in
previous work marrying continuum fermions to the lattice via an
interpolation.
For example, the counter-terms needed to maintain chiral gauge
invariance are explicitly given.
By construction coupling-constant renormalization, anomaly structure,
and (in the chiral gauge theory) fermion nonconserving amplitudes all
satisfy one's expectations from perturbative and semi-classical
analyses.
\end{abstract}
\vfill \newpage % \pagestyle{plain}

\section{Introduction}\label{intro}
A long-standing problem in quantum field theory is a nonperturbative
formulation of chiral fermions.
Our only general nonperturbative formulation of quantum field theory is
the renormalization-group limit of functional integrals defined on a
lattice.
But when chiral symmetry is an issue, there are notorious problems
\cite{Kar81,Nie81}.
Briefly, one must either sacrifice locality or positivity, tolerate
additional states (doubling or mirror states), or break chiral symmetry
explicitly.
When the coupling of fermions to gauge fields is vector-like, the
standard formulations \cite{Wil77,Sus77} are adequate, if imperfect.
On the other hand, when fermions couple to chiral gauge fields, it has
been difficult to prove a conceptually clean theory; see
ref.~\cite{Pet93} for a review.
An encouraging proposal replaces the functional integral over fermions
with an auxiliary quantum-mechanical system \cite{Nar93}, inspired by
domain-wall \cite{Kap92} and lattice Pauli-Villars \cite{Fro94}
methods.

This paper offers constructions of vector-like and chiral gauge
theories, coupling continuum fermions to lattice gauge fields by
introducing an interpolation of the latter.
Ideas of this type were first discussed by Flume and Wyler for the
Schwinger model \cite{Flu82}, and recently 't~Hooft advocated a similar
approach for four-dimensional gauge theories \cite{tHo94}.
The appeal stems from the nontrivial (instanton) topology of continuum
gauge fields, because the Atiyah-Singer index theorem \cite{Ati71}
implies an intimate relation between chiral properties of fermions and
the topology of the gauge~field.

In 1987--1988 there was some discussion about topology and fermions
in lattice gauge theory \cite{Kro87,Smi87}.
Except for a conference reports \cite{Goe92}, however, none of the
applications to chiral gauge theories have been published.
Spurred by ref.~\cite{tHo94}, I~would like to present my
variation on the theme.

Ref.~\cite{tHo94} regulates the gauge field with the lattice and the
fermions with a standard Pauli-Villars scheme.
The number of fermionic degrees of freedom (per unit volume) remains
infinite---in the words of Smit, the method is desperate \cite{Smi87}.
In particular, a numerical evaluation of the effective action would
require infinite computation, even at fixed cutoff.
This paper, on the other hand, examines a sharp cutoff on determinants,
which was first studied in ref.~\cite{And84}.
The number of fermionic degrees of freedom is now finite,
so the numerical computation of the effective action is finite~too.

Another difference between this paper's proposal and the one in
ref.~\cite{tHo94} is the strategy for removing the cutoffs.
Let $a$ denote the lattice spacing and $M$ the ultraviolet cutoff on
fermions.
In the formulation of ref.~\cite{tHo94} one takes $M\to\infty$
for $a$ fixed, and afterwards $a\to0$.
As stressed in sect.~\ref{GtH cutoff} the cutoff in ref.~\cite{tHo94}
maintains the gauge symmetry of the chiral theory only in the
$M\to\infty$ limit.
With the sharp cutoff formulated below, however, it is permissible and
natural to take $a\to0$, $M\to\infty$ with $Ma$ constant.
The latter approach is far superior in a numerical computation.
If $M$ must vary at fixed $a$, it will be extremely difficult to
generate useful ensembles of lattice gauge fields, because to obtain a
renormalized theory the bare gauge coupling must depend on $M$.

Sect.~\ref{vector-like} begins with a discussion of vector-like
theories.
The analysis starts with anomalies, because they are such a stumbling
block for lattice formulations \cite{Kar81}.
The functional integral over fermion fields is defined in the manner of
Fujikawa \cite{Fuj79}, and a specific cutoff procedure is
formulated and justified in sect.~\ref{regulator}.
These sections define an effective action for a continuous background
gauge field.
Sect.~\ref{link} summarizes the essential features of an interpolation
from the lattice field of parallel transporters to a connection.
Sect.~\ref{metric} derives a heuristic relation between the fermion
measure of this proposal and the usual measure of lattice fermions.
While inessential to the main line of argument, the derivation suggests
a rationale for relating the cutoff on determinants to the lattice
spacing.
Sect.~\ref{MC} briefly considers numerical aspects.
As usual, the generalization to chiral gauge theories is not immediate,
but sect.~\ref{chiral} produces a satisfactory definition of the chiral
effective action, including fermion nonconservation.
Finally, sect.~\ref{conclusions} remarks on some of the loose ends,
and compares the status of this formulation with ref.~\cite{Nar93}.

\section{Vector-like gauge theory}\label{vector-like}
The formal expression for the Euclidean functional integral for
fermions is
\begin{equation}\label{basic integral}
e^{-\Gamma(A)}=\int\cD\psi\cD\bar{\psi}\;e^{-S(A,\psi,\bar{\psi})}.
\end{equation}
Let us assume a background gauge potential (or connection) $A_\mu$.
In the application to lattice field theory, this connection is a
determined from the lattice gauge field, cf.~sect.~\ref{link}.
Staying momentarily ``in the continuum,'' the action is
\begin{equation}\label{vlike action}
S=\int d^4x\,\bar{\psi}(x)(\slsh{D}+m)\psi(x).
\end{equation}
Formal integration over the fermion fields yields the Boltzmann factor
\begin{equation}
e^{-\Gamma(A)}=\Det(\slsh{D}+m).
\end{equation}
The objective is to give a rigorous meaning to the measure
$\cD\psi\cD\bar{\psi}$, and/or to the determinant.

Consider the eigenfunctions and eigenvalues of the Dirac operator
$\slsh{D}$
\begin{equation}
i\slsh{D}\varphi_n=\lambda_n\varphi_n.
\end{equation}
Since $\slsh{D}$ is anti-Hermitian, the $\lambda_n$ are real.
The Dirac operator transforms covariantly under the gauge group,
so the $\lambda_n$ are gauge invariant.
The Dirac operator anti-commutes with $\gamma_5$,
i.e.\ $\gamma_5\slsh{D}=-\slsh{D}\gamma_5$.
Hence, if $\varphi_n$ is an eigenfunction with eigenvalue $\lambda_n$,
then $\gamma_5\varphi_n$ is an eigenfunction with eigenvalue
$-\lambda_n$.
As usual, one imagines that the theory is defined on a compact
space-time, and the infinite volume limit is taken at the end.
Then the spectrum of $\slsh{D}$ is discrete.

The Dirac operator can possess zero modes, $\lambda_n=0$.
In this subspace it is convenient to sort the eigenfunctions according
to chirality, i.e.\ the eigenvalue of $\gamma_5$.
Let $n_\pm$ be the number of modes with $\slsh{D}\varphi_n=0$ and
$\gamma_5\varphi_n=\pm\varphi_n$, and let $n_0=n_++n_-$.
The determinant should then be
\begin{equation}\label{unregulated}
\Det(\slsh{D}+m)=\prod_n(-i\lambda_n + m)=
m^{n_0} \prod_{\lambda_n>0} (\lambda_n^2+m^2),
\end{equation}
except that the infinite product still requires an ultraviolet
regulator; this is postponed to sect.~\ref{regulator}.
(The second equality follows because nonzero eigenvalues come in pairs
$\pm\lambda_n$.)

It is easy to see that the eigenvectors are orthonormal and form a
complete set:
\begin{equation}\label{orthonormal}
\sum_a\int d^4x\,\varphi_n^{a*}(x) \varphi_m^a(x) = \delta_{nm},
\end{equation}
\begin{equation}\label{complete}
\sum_n \varphi_n^i(x) \varphi_n^{j*}(y) = \delta(x-y)\delta^{ij},
\end{equation}
where $i$, $j$ denote spinor and color indices.
These properties of the eigenfunctions permit the expansions
\begin{equation}\label{eigenbasis}
\begin{array}{r@{\;=\;}l}
   \psi(x)    & {\displaystyle \sum_n}    a_n    \varphi_n(x), \\[1.0em]
\bar{\psi}(x) & {\displaystyle \sum_n} \bar{a}_n \varphi^\dagger_n(x),
\end{array}
\end{equation}
where the coefficients $a_n$ and $\bar{a}_n$ are Grassman numbers.
To obtain eq.~(\ref{unregulated}) from eqs.~(\ref{basic integral}) and
(\ref{vlike action}) one takes the functional integral over fields given
by eqs.~(\ref{eigenbasis}), i.e.\ the fermion measure is {\em defined\/}
to be
\begin{equation}\label{measure}
\cD\psi\cD\bar{\psi}:= \prod_n da_n d\bar{a}_n.
\end{equation}
Fujikawa \cite{Fuj79} makes a formal argument%
\footnote{Sect.~\ref{metric} pursues a similar, yet complementary, line
of thought.}
to relate the right-hand side of eq.~(\ref{measure}) to
$\prod_x d\bar{\psi}(x)d\psi(x)$.
Since the product over a continuous index is formal, it is logically
cleaner to assert eq.~(\ref{measure}) as a definition.

The Fujikawa measure is analogous to one based on Fourier modes,
\begin{equation}\label{Fourier measure}
\cD\psi\cD\bar{\psi} \sim
\prod_k d\psi(k) d\bar{\psi}(k).
\end{equation}
The momenta correspond to eigenvalues of $\slsh{\partial}$, and are also
discrete (in a box).
With a coupling to gauge fields, however, the momenta are not gauge
invariant.
Hence, the definition based on eigenfunctions of $\slsh{D}$ is
preferable.

One must check that the formalism reproduces the axial anomaly.
Consider space-time dependent chiral
transformations
\begin{equation}
\begin{array}{r@{\;=\;}l}
   \psi(x)    &        e^{-\alpha^a(x)T^a\gamma_5} \psi'(x),\\[1.0em]
\bar{\psi}(x) & \bar{\psi}'(x) e^{-\alpha^a(x)T^a\gamma_5},
\end{array}
\end{equation}
and $T^a=-(T^a)^\dagger$ is a generator of a global symmetry group.
(For U(1) take $T=i$.)
Under such transformations the measure of eq.~(\ref{measure}) is not
invariant, but
\begin{equation}
a'_n= \int d^4x\,
\varphi^\dagger_n(x)e^{\alpha^a(x)T^a\gamma_5}\varphi_m(x)\,a_m
=: C_{nm} a_m,
\end{equation}
and similarly for $\bar{a}'_n$, with sums on $m$ implied.
The rules of Berezin integration imply
\begin{equation}
\prod_n da_n=\Det C \prod_n da_n'.
\end{equation}
The Jacobian determinants are then responsible for the anomaly.
Using an eigenfunction expansion of $\alpha^a(x)T^a\gamma_5$ one can
write $C_{nm}=\exp c_{nm}$ with
\begin{equation}
c_{nm}=
\int d^4x\,\varphi^\dagger_n(x)\gamma_5T^a\varphi_m(x)\alpha^a(x).
\end{equation}
Then
\begin{equation}\label{detC}
\Det C = e^{\Tr c} =
\exp\left(\int d^4x\,\cA^a(x)\alpha^a(x) \right),
\end{equation}
where the anomaly
\begin{equation}\label{anomaly}
\cA^a(x) = \sum_n \varphi^\dagger_n(x)\gamma_5T^a\varphi_n(x).
\end{equation}
Collecting the anomaly from $\cD\bar{\psi}$ and $\cD\psi$, as well as
terms from the chiral transformation of the action, yields the
(anomalous) Ward-Takahashi identity
\begin{equation}\label{aWTi}
\partial_\mu\bar{\psi}(x)\gamma_\mu\gamma_5T^a\psi(x)=
\bar{\psi}(x)\{T^a,m\}\gamma_5\psi(x)+2\cA^a(x).
\end{equation}

\section{Regulating the fermions}\label{regulator}
\begin{figure}[b]
\epsfxsize=\textwidth \epsfbox{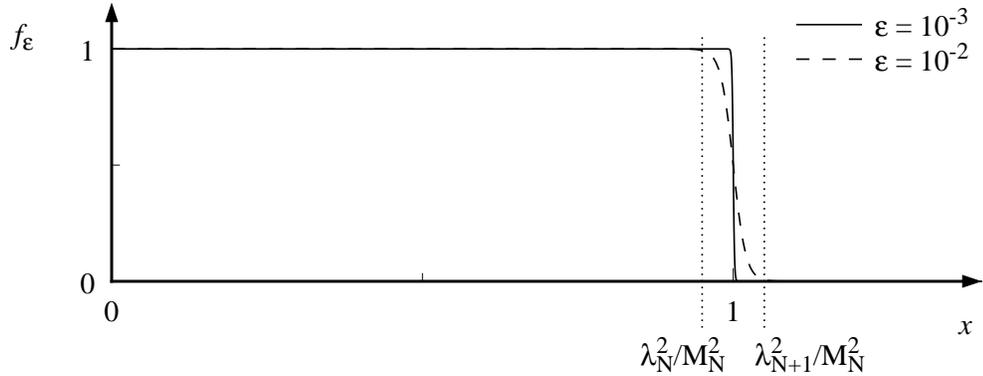}
\caption[regulator-function]{The shape of the regulator function
$f_\varepsilon(x)$, which is designed so that modes with eigenvalues
$\lambda_n^2>\lambda_N^2$ are omitted.}\label{regulator-function}
\end{figure}

The preceding discussion skirts the need to regulate the determinants
in the ultraviolet.
Let us start with a kind of Pauli-Villars regulator.
Eq.~(\ref{anomaly}) becomes
\begin{equation}\label{regulated anomaly}
\cA^a_{\rm reg}(x) = \sum_n f_{\varepsilon_N}(\lambda_n^2/M_N^2)
\varphi^\dagger_n(x)\gamma_5T^a\varphi_n(x).
\end{equation}
Let the index $n$ run over $\{1-n_0,\ldots,0, 1, 2,\ldots\}$ with
the convention that $n>0$ denotes nonzero modes; let us order the
nonzero modes by $\lambda_n^2$ and take $n$ odd (even) if $\lambda_n$
is negative (positive).
The regulating function $f_\varepsilon(x)$ is chosen to look like the
sketch in fig.~\ref{regulator-function}.
The cutoff $M_N$ and smearing parameter $\varepsilon_N$ should satisfy
\begin{equation}
|\lambda_N|<M_N<|\lambda_{N+1}|,
\end{equation}
\begin{equation}
\varepsilon_N\ll(\lambda_{N+1}^2-\lambda_N^2)/M_N^2,
\end{equation}
where $N$ is even (so $\lambda_{N+1}\neq-\lambda_N$), and one takes
$\varepsilon_N\to0$ with $M_N$ fixed.%
\footnote{The notation ignores accidentally degenerate nonzero modes,
but it is clear what to do.}
The Fermi function
\begin{equation}
f_\varepsilon(x)=\frac{e^{-1/\varepsilon}+1}{e^{(x-1)/\varepsilon}+1}
\end{equation}
changes rapidly from 1 to 0 as $\real x$ passes through 1.
Since $f_\varepsilon(0)=1$ and $f_\varepsilon$ and all its derivatives
vanish at $\infty$, it supports the standard analysis \cite{Fuj79}
that leads to
\begin{equation}\label{FdualF}
\cA^a_{\rm reg}=\frac{\tr T^a}{16\pi^2}\tr[F_{\mu\nu}{}^*F_{\mu\nu}]
\end{equation}
in four dimensions, where the two traces are over global and gauge-group
indices.
Details in Appendix~\ref{convergence} show that all higher dimension
terms, e.g.\ $\tr(F^{n+2})/M^{2n}$, are proportional
to $\varepsilon_N^{-n}e^{-1/\varepsilon_N}$ as $\varepsilon_N\to0$.

To check that the regulator does not spoil the derivation of
eq.~(\ref{aWTi}), one restricts the chiral transformations $\alpha(x)$
to those for which
\begin{equation}\label{block c}
\int d^4x\,\varphi^\dagger_n(x)\alpha^a(x)T^a\gamma_5\varphi_m(x)=0,
\end{equation}
if one of $n,~m>N$.
If both $n$ and $m>N$, then the left-hand side of eq.~(\ref{block c})
need not vanish.
The Jacobian matrix $C$ takes a block form, and the regulator decouples
the block with large eigenvalues.
Nevertheless, the transformation function $\alpha(x)$ is sufficiently
arbitrary to derive the Ward-Takahashi identity.

The $\varepsilon_N\to0$ limit corresponds to keeping the first $N$ modes
of $\slsh{D}$ completely and eliminating the rest.
In a loose analogy with the lattice cutoff, in which the Fourier measure
of eq.~(\ref{Fourier measure}) is truncated, one can carry out
the truncation of the modes throughout.
Thus, the regulated measure is
\begin{equation}\label{regulated measure}
\left(\cD\psi\cD\bar{\psi}\right)_N =
\prod_{n=1-n_0}^N da_n d\bar{a}_n,
\end{equation}
with eigenvalues ordered as above.
The functional integral is now over fields given by
\begin{equation}
\begin{array}{r@{\;=\;}l}
   \psi(x)    &
{\displaystyle \sum_{n=1-n_0}^N}    a_n   \varphi_n(x),\\[1.0em]
\bar{\psi}(x) &
{\displaystyle \sum_{n=1-n_0}^N} \bar{a}_n\varphi_n^\dagger(x).
\end{array}
\end{equation}
The effective Boltzmann factor becomes
\begin{equation}\label{regulated effective action}
e^{-\Gamma_N(A)}= \prod_{n=1-n_0}^N(-i\lambda_n+m)=
m^{n_0}
\prod_{\scriptsize\begin{array}{c}n=1\\ \lambda_n>0\end{array}}^{N}
(\lambda_n^2+m^2).
\end{equation}
Similarly, the regulated fermion propagator is
\begin{equation}
\int\left(\cD\psi\cD\bar{\psi}\right)_N \psi(x) \bar{\psi}(y) e^{-S} =
\sum_{n=1-n_0}^N
\varphi_n(x) \frac{1}{-i\lambda_n + m} \varphi_n^\dagger(y)
\,e^{-\Gamma_N(A)}.
\end{equation}
The finite-mode cutoff has been examined before \cite{And84},
with emphasis on anomalies.
Those papers did not notice the disappearance of higher-dimensional
terms as the Fermi function becomes infinitely sharp, a feature that is
especially important in the chiral gauge theory, cf.~sect.~\ref{chiral}.

In perturbation theory, one must be careful to impose the sharp cutoff
with $\slsh{D}$, rather than with loop momenta.
For example, the dependence of the regulator on $\slsh{D}$, and hence
on $A_\mu$, induces additional terms in the gauge current, to maintain
gauge invariance.
%As a consequence, the Feynman rules for this and other vertices are
%similar to those for Wilson fermions.

How should one choose $N$?
One ought to take nonzero modes in pairs: it would be silly to take
$\lambda_n$ and not $-\lambda_n$.
One also ought to retain all zero modes, because they are the most
infrared of all!
But from the Atiyah-Singer index theorem \cite{Ati71} the number of zero
modes is even (odd) if the topological charge $Q$ is even (odd).
Thus, $N$ must depend on the gauge field and be even (odd) if $Q$ is
even (odd).

The order of magnitude of $N$ can be specified only vaguely.
An examination of cutoff effects, Appendices~\ref{convergence}
and~\ref{cutoff}, introduces expansions.
For a generic cutoff, one would need to maintain $A_\mu/M\ll1$.
The absence of non-universal terms with the finite-mode cutoff, however,
permits a lower cutoff.
If, as in the next section, $A_\mu$ is obtained from a lattice gauge
field, $|A|<C/a$, where $C$ is a gauge-group dependent constant.
With the finite-mode cutoff it is thus natural to take $M_N\sim a^{-1}$.
Indeed, on an $N_S^3\times N_T$ lattice one expects
\begin{equation}
N_{\rm lat}=4RN_S^3N_T
\end{equation}
fermionic degrees of freedom in the lattice fermion field (per flavor),
where $R$ is the dimension of the fermion's gauge-group representation.
An obvious choice would be to augment the usual number with the zero
modes, i.e.~$N=n_0+N_{\rm lat}$.

\section{Coupling to a lattice gauge field}\label{link}
To extend the formulation to lattice gauge theory, one must provide an
interpolation determining a connection $A_\mu(x)$ from the lattice gauge
field $U_\mu(x)$.
Ref.~\cite{tHo94} gives a prescription adequate for defining traces
of products of $F_{\mu\nu}$, but does not prescribe $A_\mu$.
Other possibilities, based on definitions of the topological (instanton)
charge of a lattice gauge field \cite{Lue82,Phi86}, do prescribe $A_\mu$
explicitly \cite{Goe93,Kro88}.
The latter have two crucial properties:
\begin{enumerate}
\item
The only singularities in $A_\mu$ are instanton-like.
\item
The interpolation $A_\mu$ transforms as a connection under
{\em lattice\/} gauge transformations.
\end{enumerate}
Ref.~\cite{tHo94} does prove a useful theorem on the spectrum of the
Dirac operator for a bounded gauge field.
The (manifestly) topological interpolations obey the hypothesis of the
theorem.

The great insight of ref.~\cite{Lue82} was to recognize that the lattice
gauge field can be used to define a fiber bundle.
Then the topological charge is the second Chern number of the bundle.
L\"uscher originally provided expressions for transition functions,
which encode changes of gauge
\begin{equation}
A_\mu^{(\alpha)}=
v_{\alpha\beta}(\partial_\mu+A_\mu^{(\beta)})v_{\beta\alpha}
\end{equation}
from a patch $\alpha$ of space-time to a neighboring one $\beta$.
For consistency, $v_{\beta\alpha}=v_{\alpha\beta}^{-1}$.
In the fiber-bundle formalism, the gauge is fixed separately within each
patch, such that $A_\mu^{(\alpha)}$ has no singularities.
The winding responsible for topological charge then resides in the
transition functions \cite{Lue82,vBa82}.
The more familiar patch-independent connection is
\begin{equation}\label{connection}
A_\mu = w_\alpha^{-1}(x)(\partial_\mu+A_\mu^{(\alpha)}) w_\alpha(x),
\end{equation}
where the section $w_\alpha$ is related to the transition functions by
\begin{equation}\label{v=ww}
v_{\alpha\beta}=w_\alpha w_\beta^{-1}.
\end{equation}
The first step of refs.~\cite{Goe93,Kro88} is a patch-wise continuous,
bounded interpolation for the nonsingular $A_\mu^{(\alpha)}$.
Eq.~(\ref{connection}) produces a continuous, bounded connection
$A_\mu$, but when the transition functions have nontrivial winding,
the sections, and thus $A_\mu$, have directional singularities.
By construction, therefore, the {\em only\/} singularities in the
globally defined $A_\mu$ are those induced by the instanton-winding of
the section.

The other crucial property of the reconstructed gauge potentials of
refs.~\cite{Goe93,Kro88} is the response to a lattice gauge
transformation.
The transformation law of the lattice gauge field is
\begin{equation}
{}^gU_\mu(s)= g(s) U_\mu(s)g^{-1}(s+\hat{\mu}),
\end{equation}
where $s$, $s+\hat{\mu}$ denote lattice sites.
For $x$ in the patch $\sigma$ associated with $s$,
the reconstructed section obeys \cite{Goe87,Kro88}
\begin{equation}
{}^gw_{\sigma}(x) = g(s) w_{\sigma}(x)g^{-1}(x).
\end{equation}
The interpolation $g(x)$ is independent%
\footnote{If $x\in\sigma\cap\tau$, eq.~(\ref{v=ww}) requires that
the interpolation obey $g(x)|_\sigma=g(x)|_\tau$.}
of $\sigma$, but, of course, $g(s)$ is independent of $x$.
The interpolated connection transforms as
% \begin{equation}
% {}^gA_\mu^{(\sigma)} = g(s) A_\mu^{(\sigma)} g^{-1}(s)
% \end{equation}
\begin{equation}\label{transform A}
{}^gA_\mu = g(x)(\partial_\mu+A_\mu) g^{-1}(x),
\end{equation}
with the {\em same\/} function $g(x)$.

If $g(s)=g_2(s)g_1(s)$ at each site, the three interpolated gauge
transformation fields obey the composition law
\begin{equation}\label{composition}
g(x;U)= g_2(x;{}^{g_1}U) g_1(x;U).
\end{equation}
Note that the interpolation depends on the underlying lattice gauge
field, which is emphasized here by the second argument.
Consequently, every interpolated $g(x;U)$ can be built up from
infinitesimal, site-by-site steps.

For the present discussion the complicated expressions for the
interpolated $g(x)$ and $A_\mu$ are not illuminating.
Interested readers can consult refs.~\cite{Goe87,Goe93}
for L\"uscher's bundle.
(For Phillips and Stone's bundle, analogous results can be obtained
\cite{Kro88}.)

Our prescription for coupling fermions to lattice gauge fields is to
start with the lattice gauge field $U_\mu$, interpolate to obtain the
connection $A_\mu$, use the associated Dirac operator to define the
measure, and regulate the determinant using the sharp cutoff.
Because the Dirac operator is constructed using fiber bundles, it
automatically satisfies the Atiyah-Singer index theorem
\begin{equation}
n_+-n_-=Q,
\end{equation}
where $Q$ is the topological charge (as defined by ref.~\cite{Lue82})
of the lattice gauge field.
Moreover, from eq.~(\ref{transform A}) the eigenvalues of the Dirac
operator, and hence the effective action, are invariant under lattice
gauge transformations.
  From eq.~(\ref{regulated effective action}) the Boltzmann factor
$e^{-\Gamma_N(U)}$ is also positive and finite.

\section{Relation to lattice fermion
fields\protect\footnotemark}\label{metric}
\footnotetext{This section is a diversion, and the rest of the paper
makes no use of it.}
It is intriguing to contrast the regulated measure
of eq.~(\ref{regulated measure}) with the usual one
\begin{equation}
\left(\cD\psi\cD\bar{\psi}\right)_{\rm lat}=
\prod_s d\psi(s) d\bar{\psi}(s),
\end{equation}
where $\psi(s)$ and $\bar{\psi}(s)$ denote the lattice fermion
field and its conjugate at site $s$.
There is an interpolation of the fermion field analogous to $A_\mu$.
In particular, under a lattice gauge transformation the interpolated
fermion field transforms as
\begin{equation}
{}^g\psi(x)=g(x)\psi(x),\quad
{}^g\bar{\psi}(x)=\bar{\psi}(x)g^{-1}(x),
\end{equation}
with the same interpolated $g(x)$ mentioned in sect.~\ref{link}.

Let us expand the interpolated field $\psi(x)$ as in
eq.~(\ref{eigenbasis}), use the orthonormality to solve for the $a_n$
and take differentials.
Then
\begin{equation}\label{a-psi}
\begin{array}{r@{\;=\;}l}
da_n & {\displaystyle\sum_s\int}d^4x\,\varphi^\dagger_n(x)
\dfrac{\partial\psi(x)}{\partial\psi(s)}d\psi(s)
=: {\displaystyle\sum_s} m_{ns}d\psi(s), \\[1.2em]
d\bar{a}_n & {\displaystyle\sum_s\int}d^4x\,d\bar{\psi}(s)
\dfrac{\partial\bar{\psi}(x)}{\partial\bar{\psi}(s)} \varphi_n(x)
=: {\displaystyle\sum_s} d\bar{\psi}(s)\bar{m}_{sn}.
\end{array}
\end{equation}
If the number of modes $N$ equals the number of lattice sites,
$m$ and $\bar{m}$ are $N\times N$ matrices and one can write
\begin{equation}
\prod_n da_nd\bar{a}_n=
({\det}^{-1}m \times {\det}^{-1}\bar{m})\prod_sd\psi(s) d\bar{\psi}(s).
\end{equation}
The determinants here are akin to ones appearing in ref.~\cite{Fuj79}.
Formally, one can combine the determinants into a ``metric''
\begin{equation}
g_{\bar{s}s}=\bar{m}_{\bar{s}n} m_{ns} =
\int d^4x\,\frac{\partial\bar{\psi}(x)}{\partial\bar{\psi}(\bar{s})}
\frac{\partial\psi(x)}{\partial\psi(s)},
\end{equation}
and rewrite the measure as
\begin{equation}\label{metric measure}
\prod_n^N d\bar{a}_n da_n=
\det g^{-1} \prod_s d\psi(s)d\bar{\psi}(s).
\end{equation}
The metric depends covariantly on the gauge field, by construction of
the interpolation.

One can develop a geometric picture by imagining a Grassman line element
\begin{equation}
\sum_{\bar{s}s}g_{\bar{s}s}d\bar{\psi}(\bar{s}) d\psi(s)=
\sum_n d\bar{a}_nda_n
\end{equation}
that is gauge invariant.
In this language the customary lattice fields are curvilinear
coordinates, and the expansion coefficients are rectilinear.
According to this picture, the usual lattice measure mistakenly neglects
the curvature.
Note that without the gauge interaction, the metric becomes flat:
the two bases are then Fourier transforms of one another.

The metric also becomes trivial in the naive continuum limit.
Then the interpolation is unnecessary and
$\partial\psi(x)/\partial\psi(s)=\delta(x-s)$.
The factor $\det g^{-1}$ is essential, however, for obtaining the
correct anomaly, index theorem, etc.
In particular, the anomaly arises because one may remove the regulators
only after calculating with the functional integral, not (as in the
naive continuum limit) before.

Of course, eq.~(\ref{metric measure}) is merely heuristic.
To make the manipulations rigorous, the interpolated fields must be
smooth enough that the Dirac eigenmode expansions leading to
eq.~(\ref{a-psi}) stop at the $N$th term.
Here $N$ is both the number of modes kept and the number of lattice
sites.
This condition presumably puts constraints on the smoothness of the
lattice gauge field similar to, but perhaps more stringent than,
those imposed by uniqueness considerations in the fiber bundle
constructions \cite{Lue82,Phi86}.
One should emphasize, however, that such constraints are on the
connection between the present measure and the usual lattice one.
They do not detract from eq.~(\ref{measure}) as a definition of the
functional integral, or $e^{-\Gamma_N(U)}$ as a definition of the
effective action.

\section{Computational Considerations}\label{MC}
Since the effective action $\Gamma_N(U)$ is real and positive, the
functional integral over the gauge bosons is amenable to the Monte Carlo
method with importance sampling.
The Dirac eigenvalues, and hence the effective action, depend on the
lattice gauge field $U$ so they must be recomputed for every change
of~$U$.
If the effective action is defined with a Gaussian
$e^{-\lambda_n^2/M^2}$ \cite{Fuj79} or standard Pauli-Villars
\cite{tHo94} regulator, one would need to compute all eigenvalues of
$\slsh{D}$ and weight them appropriately---an infinite computation.
With eq.~(\ref{regulated effective action}), however, only the lowest
$N$ eigenvalues are needed---a finite computation.%
\footnote{Conceding round-off error, the ``infinite'' computation is
$\epsilon^{-1}$ times more costly, where $\epsilon$ characterizes
machine precision.}

To compute the eigenvalues one can introduce an auxiliary lattice much
finer than the original one.
Parallel transporters for the fine lattice are constructed from the
interpolated gauge field.
Now consider any discretization of the Dirac operator, and denote its
eigenvalues by $d_n$.
The discretization and auxiliary lattice spacing must be chosen such
that $d_n=\lambda_n$ (up to tolerable floating-point precision) for
$n\leq N$.
On a fine enough auxiliary lattice, ``any'' discretization becomes
precise enough.
To avoid problems sorting the eigenvalues, however, a discretized
operator without a doubled spectrum is preferable.
Suitable examples would be the Wilson discretization \cite{Wil77},
or one derived by gauging the fixed-point action of a free fermion
\cite{Wie93}.
The discretization may break chiral symmetry, provided the breaking
is numerically significant only for modes above the cutoff~$N$.

Similar remarks apply to the construction for chiral gauge theories,
sect.~\ref{chiral}.
An important difference is that the Boltzmann factor $e^{-\Gamma_N(U)}$
can be complex (for fermions in ``complex'' representations), and in any
case not positive definite.
Monte Carlo integration is then much more difficult, because
fluctuations in sign reduce the effectiveness of importance sampling,
cf.~sect.~\ref{how to}.
Nevertheless, $e^{-\Gamma_N(U)}$ can be evaluated with finite
computation.

\section{Chiral fermions}\label{chiral}
\subsection{General remarks}
The preceding sections provide a definition of the functional integral
for vector-like fermions.
It is nonperturbative, has the correct axial anomaly, and by
construction provides a natural association between a Dirac operator
and a topological charge, so that the index theorem is obeyed.
The ideas are now extended to chiral fermions, which appear in the
Standard Model and in grand unified theories.

The essential feature of chiral gauge theories is that positive and
negative chirality fermions transform under different representations
of the gauge group.
In a basis of the Dirac matrices with $\gamma_5$ diagonal it is useful
to split the four-component Dirac spinor into two two-component Weyl
spinors.
Without loss of generality one can charge-conjugate the negative
chirality part and assemble everything into one positive chirality
field.
This Weyl spinor is henceforth denoted $\psi_+$, and the representation
of the gauge group under which it transforms is denoted $\rho$,
with generators~$t^a$.
The kinetic term of the action is
\begin{equation}\label{chiral action}
S=\int d^4x\,\psi^\dagger_+\slsh{D}_+\psi_+ .
\end{equation}
As before one wants to define the Boltzmann factor
\begin{equation}\label{chiral Gamma}
e^{-\Gamma(A)}=\int\cD\psi_+\cD\psi_+^\dagger
e^{-S(A,\psi_+,\psi^\dagger_+)}.
\end{equation}
But $\slsh{D}_+$ maps positive chirality Weyl spinors into negative
chirality Weyl spinors.
The underlying difficulty in constructing a chiral gauge theory is that
the right-hand side of eq.~(\ref{chiral Gamma}) is {\em not\/} a
(functional) determinant.

A well-formulated chiral gauge theory should also exhibit fermion
nonconservation \cite{tHo76}.
If the vector-like operator $\slsh{D}$ has zero modes, the right-hand
side of eq.~(\ref{chiral Gamma}) should vanish:
Recall that the zero modes have definite chirality.
Thus the zero modes possess natural projections from Dirac spinors
$\varphi_\pm$ onto Weyl spinors $\phi_\pm$.
For positive chirality the projection implies $\slsh{D}_+\phi_+=0$,
and for negative chirality $\phi_-^\dagger\slsh{D}_+=0$
(integration by parts implied).
In eq.~(\ref{chiral Gamma}) integration over the $\phi_+$ component of
$\psi_+$ or over the $\phi_-^\dagger$ component of $\psi_+^\dagger$
yields zero.
On the other hand, if there are $n_\pm$ zero modes of each chirality
(counting individual species appropriately) the integral
\begin{equation}\label{B violation}
\cZ(\nu_+,\nu_-)=
\int\cD\psi_+\cD\psi_+^\dagger\,(\psi_+^\dagger)^{\nu_-}(\psi_+)^{\nu_+}
e^{-S(A,\psi_+,\psi^\dagger_+)}
\end{equation}
does not vanish if $\nu_\pm\ge n_\pm$.
(The notation $(\psi_+)^{\nu_+}$ is schematic for a product of $n_+$
suitable components or positions of $\psi_+$.)
Amplitudes for fermion nonconservation are proportional to integrals
like $\cZ(\nu_+,\nu_-)$.

Before discussing how to regulate the integrals in
eqs.~(\ref{chiral Gamma}) and (\ref{B violation}), one should list
the properties of eq.~(\ref{chiral Gamma}) that the regulator should
respect.
In addition to the connection between zero modes and fermion
nonconservation, one wants
\begin{enumerate}
\item \label{invariant}
$\real\Gamma(^gA)=\real\Gamma(A)$, under all circumstances.
\item \label{anomaly free?}
$\imag\Gamma(^gA)=\imag\Gamma(A)$, only if $\rho$ is ``anomaly-free.''
\item \label{complex}
$\imag\Gamma(A)\neq0$, if $\rho$ is complex;%
\footnote{If there is a unitary matrix $u$ such that
$t^a=ut^{a*}u^\dagger$, then $\rho$ is real; otherwise it is complex.
The representation generated by $t^{a*}$ is denoted $\rho^*$.
If $\rho$ is real, then $e^{-\Gamma(A)}$ is real.}
indeed $\imag\Gamma(\rho^*(A))=-\imag\Gamma(\rho(A))$.
\end{enumerate}
Here $^gA_\mu=g(\partial_\mu+A_\mu)g^{-1}$ is a gauge transform of
$A_\mu$, and the notation $\Gamma(\rho(A))$ stresses the fermion
representation.
Condition~\ref{anomaly free?} is imposed so that the nonperturbative
definition reproduces perturbation theory.
Condition~\ref{complex} identifies a diagnostic feature of the effective
action of chiral fermions.

Even though the right-hand side of eq.~(\ref{chiral Gamma}) is by nature
not a determinant, ultraviolet regulators almost always turn it into
one.
For example, in the most naive (and unsuccessful) lattice formulation,
$\slsh{D}_+$ becomes a large, square, numerical matrix, and
$\det\slsh{D}_+$ is the matrix determinant.
So let us provisionally assume
\begin{equation}
e^{-\Gamma(A)}=\prod_n^N(-i\lambda_n).
\end{equation}
In anticipation of a cutoff analogous to the one for vector-like
theories (sect.~\ref{regulator}), the product runs only over the first
$N$ eigenvalues.
The most transparent way to realize Conditions~\ref{invariant}
and~\ref{anomaly free?} is as follows:
Write $\lambda_n=ie^{l_n+i\theta_n}$, with $l_n$ and $\theta_n$ real,
and note that
\begin{equation}\label{Gamma as det}
\real\Gamma(A)=-\sum_n^Nl_n,\quad
\imag\Gamma(A)=-\sum_n^N\theta_n~\bmod 2\pi.
\end{equation}
If the moduli $|\lambda_n|=e^{l_n}$ of the eigenvalues are gauge
invariant and one orders the eigenvalues by $|\lambda_n|$,
Condition~\ref{invariant} is satisfied.
By Condition~\ref{anomaly free?}, in anomalous (sub)representations the
phases of the eigenvalues would be gauge variant, but the variation from
one species could cancel that of another.
Unfortunately, it seems that eigenvalue problems with these simple
gauge-transformation properties leave the total phase $\imag\Gamma$
unspecified, flouting Condition~\ref{complex}.%
\footnote{Ref.~\cite{Goe92} adopts the spirit of this realization.
There $\real\Gamma$ is related to the vector-like theory and
$\imag\Gamma=:\eta$ to the spectral asymmetry of a certain operator
\cite{Alv86}.
This method, however, requires three regulators: one for the gauge
fields, one for the vector-like fermions, and one for $\eta$.}

\subsection{A specific formulation}
A standard way to cast the effective action as a determinant is to
introduce a new negative chirality partner $\psi_-$ with no dynamics.
The action is now
\begin{equation}\label{extended chiral action}
S=\int d^4x\,\left(\psi^\dagger_+\slsh{D}_+\psi_+ +
\psi^\dagger_-\slsh{\partial}_-\psi_- \right)
=\int d^4x\,\bar{\psi}\hat{D}\psi,
\end{equation}
where the four-component spinor
\begin{equation} \label{Dirac spinor}
\psi= \left(\begin{array}{c} \psi_+ \\ \psi_- \end{array}\right),
\quad\bar{\psi}= \left(\psi_-^\dagger \;\; \psi_+^\dagger\right),
\end{equation}
and
\begin{equation}\label{D hat}
\hat{D}:=\slsh{D}_++\slsh{\partial}_-=
\left(\begin{array}{cc} 0 & \slsh{\partial} \\
\slsh{D} & 0 \end{array}\right).
\end{equation}
The matrix forms of eqs.~(\ref{Dirac spinor}) and (\ref{D hat}) presumes
a $\gamma$-matrix basis with $\gamma_5$ diagonal.
Formally, the functional integral over $\cD\psi_-\cD\psi_-^\dagger$ is
trivial, but the combined integral
\begin{equation}
e^{-\tilde{\Gamma}}=\int\cD\psi\cD\bar{\psi}\,e^{-S(\psi,\bar{\psi},A)}
\end{equation}
can be expressed as a determinant, as in the vector-like theory.

The operator $i\hat{D}$ is not self-adjoint, but it is elliptic
\cite{Alv84}, so on a compact space-time it still has a discrete
spectrum.
The eigenvalue problem now has different right and left eigenfunctions.
There are zero modes
\begin{equation}
 i\hat{D}         \varphi_{+,n} = 0, \quad
(i\hat{D})^\dagger\varphi_{-,n} = 0,
\end{equation}
where $\varphi_{\pm,n}$ are zero-mode eigenfunctions of the
vector-like operator $\slsh{D}$, with chirality $\pm1$.
Nonzero modes come in pairs%
\footnote{The eigenvalue problem described here is implicitly adopted by
ref.~\cite{tHo94} and many other papers \cite{And84,Bal82,Ein84}.
Often the literature discusses models that couple fermions to external
fields via $\slsh{\partial}+\slsh{V}+\slsh{A}\gamma_5$.
This operator is again not self-adjoint, though elliptic, so there
are left and right eigenfunctions $\chi_n\neq\eta_n$.
Most papers either ignore this subtlety, or try to circumvent it.
For example, some tricks turn $i\hat{D}$ into a Hermitian operator,
thus leaving $\imag\Gamma$ unspecified.
They cannot be adopted here.}
\begin{equation}
\begin{array}{r@{\;=\;}l}
 i\hat{D}         \eta_n & \lambda_n   \eta_n, \\[1.0em]
(i\hat{D})^\dagger\chi_n & \lambda_n^* \chi_n
\end{array}
\end{equation}
that are mutually orthonormal:
\begin{equation}
\int d^4x\,\chi_n^{\dagger}(x) \eta_m(x) = \delta_{nm}.
\end{equation}
In addition
$\int d^4x\,\chi_n^\dagger       \varphi_{+,m}=
 \int d^4x\,\varphi_{-,n}^\dagger\eta_m=
 \int d^4x\,\varphi_{-,n}^\dagger\varphi_{+,m}=0$.
Furthermore, since $\gamma_5\hat{D}=-\hat{D}\gamma_5$, if $\eta_n$ has
eigenvalue $\lambda_n$, then $\gamma_5\eta_n$ has eigenvalue
$-\lambda_n$, and similarly for $\chi_n$, $\gamma_5\chi_n$.

The functional integral is now defined to be over fields
\begin{equation}\label{chiral fields}
\begin{array}{r@{\;=\;}l@{\;+\;}l}
   \psi(x)    &
{\displaystyle \sum_{n=1}^{n_+}}   z_n    \varphi_{+,n}(x) &
{\displaystyle \sum_{n=1}^N}       a_n    \eta_n(x), \\[1.0em]
\bar{\psi}(x) &
{\displaystyle \sum_{n=1}^{n_-}}\bar{z}_n \varphi_{-,n}(x) &
{\displaystyle \sum_{n=1}^N}    \bar{a}_n \chi^\dagger_n(x),
\end{array}
\end{equation}
i.e.\ the measure is
\begin{equation}\label{chiral measure}
\cD\psi_-\cD\psi_-^\dagger\cD\psi_+\cD\psi_+^\dagger=
\cD\psi\cD\bar{\psi}=
\prod_{n=1}^N da_n d\bar{a}_n
\prod_{n=1}^{n_+} dz_n \prod_{n=1}^{n_-} d\bar{z}_n.
\end{equation}
As in sect.~\ref{regulator} the functional integral is cut off by
retaining only the lowest $N$ nonzero modes.
The principle for ordering the eigenvalues is revealed below, but
clearly the two modes with the same $\lambda_n^2$ should be
adjacent.
Analogously to sect.~\ref{regulator}, it is convenient to take $n$ odd
(even) if $\real\lambda_n$ is negative (positive).

Let us first integrate over the $a_n$ and $\bar{a}_n$.
Eqs.~(\ref{chiral fields}) and (\ref{chiral measure}) yield
\begin{equation}\label{Gamma tilde}
e^{-\tilde{\Gamma}_N(A)}=\prod_{n=1}^N(-i\lambda_n)=
\prod_{\scriptsize\begin{array}{c}n=1\\ \real\lambda_n>0\end{array}}^{N}
\lambda_n^2.
\end{equation}
for the functional integral.
In a real representation these eigenvalues come in pairs
$(-i\lambda_n,i\lambda_n^*)$;
$\tilde{\Gamma}_N$ satisfies Condition~\ref{complex}.
But one should not expect $\tilde{\Gamma}_N(A)$ to be a suitable
definition of the effective action---hence the tilde---because neither
the moduli nor the phases of the nonzero eigenvalues are gauge
invariant.

Under the gauge transformation $g=e^\omega$ the chiral Dirac operator
transforms as $^g\hat{D}=e^{\omega P_-}\hat{D}e^{-\omega P_+}$, where
$P_\pm=\half(1\pm\gamma_5)$.
The zero modes are gauge invariant.
To first order in $\omega$ the nonzero eigenvalues vary by
\begin{equation}
{}^g\lambda_n = \lambda_n \left(1 -
\int d^4x\,\chi_n^\dagger(x)t^a\gamma_5\eta_n(x)\omega^a(x)\right),
\end{equation}
which immediately yields the variation of $\tilde{\Gamma}_N$:
\begin{equation}\label{gauge variation}
\delta_\omega\tilde{\Gamma}_N=\int d^4x\,\cA^a_{\rm reg}(x)\omega^a(x),
\end{equation}
where
\begin{equation}\label{chiral anomaly}
\cA^a_{\rm reg}(x)=\sum_{n=1}^N \chi_n^\dagger(x)t^a\gamma_5\eta_n(x)
=\lim_{\varepsilon_N\to0}\sum_{n=1}^\infty \chi_n^\dagger(x)t^a\gamma_5
f_{\varepsilon_N}(-\hat{D}^2/M_N^2)\eta_n(x).
\end{equation}
The last expression applies if the eigenvalues are ordered by increasing
$\real\lambda_n^2$, which is justified because it reproduces the
consistent gauge anomaly.

  From eq.~(\ref{chiral anomaly}) and Appendix~\ref{convergence},
the imaginary part of $\delta_\omega\tilde{\Gamma}_N$ is
\begin{equation}\label{consistent anomaly}
i\imag\cA^a_{\rm reg}=
\frac{1}{24\pi^2}\varepsilon_{\mu\nu\sigma\tau}
\partial_\mu{\tr}_\rho[t^a(A_\nu\partial_\sigma A_\tau +
\half A_\nu A_\sigma A_\tau )],
\end{equation}
the familiar consistent anomaly \cite{Bar69,Wes71,Gro72}.
It vanishes if $\tr_\rho(t^a\{t^b,t^c\})=0$ in representation $\rho$
\cite{Gro72};
$\tilde{\Gamma}_N$ satisfies Condition~\ref{anomaly free?}.

Even if the anomaly cancels, however, the real part
$\delta_\omega\tilde{\Gamma}_N$ is not gauge invariant.
With $\alpha_2^a$ and $\alpha_{4R}^a$ from Appendix~\ref{convergence}
\begin{equation}
\delta_\omega\real\tilde{\Gamma}_N=
\frac{1}{16\pi^2}\int d^4x\,
\left(-M_N^2\alpha_2^a(x)+\alpha_{4R}^a(x)\right)\omega^a(x).
\end{equation}
After taking $\varepsilon_N\to0$, all higher-dimension terms drop out
because they are proportional to $e^{-1/\varepsilon_N}$.
Following Bardeen, the gauge variation of the real part
can be compensated by counter-terms \cite{Bar69}.
Let
\begin{equation}\label{counter-terms}
\begin{array}{r@{\,=\,}l}
S_2 & -
{\displaystyle \frac{M_N^2}{16\pi^2}\int}d^4x\,\tr_\rho( A^2 ),
 \\[1.0em]
S_4 &
{\displaystyle \frac{1}{48\pi^2}\int}d^4x\,\tr_\rho\left[
\half  A_\mu\partial^2A_\mu    + (\partial\cdot A)^2 +
\quart A_\mu A_\nu A_\mu A_\nu - \half (A^2)^2 \right],
\end{array}
\end{equation}
and $S_{\rm ct}=S_2+S_4$.
If $S_{\rm ct}$ is computed using the {\em interpolated\/} gauge field,
its gauge variation cancels that of $\real\tilde{\Gamma}_N$ exactly.%
\footnote{In other methods \cite{Bor90,Alo90,Bod91} the cancellation is
either approximate or subject to tuning.}
The choice of $S_4$ is not unique, because one could also add a gauge
invariant term proportional to $\tr_\rho F^2$.
But the ambiguity corresponds to a shift in the (inverse) bare gauge
coupling, so it should make no difference once all cutoffs are removed.

The counter-term $S_2$ is supposed to remove a quadratic divergence,
but the number of modes $N$---not the mass $M_N$ that appears in
eq.~(\ref{counter-terms})---defines the cutoff.
The number of fermion modes with momentum below a cutoff $M$ is
\begin{equation}
N=4R\sum_k f_\varepsilon(k^2/M^2).
\end{equation}
Approximating the sum by an integral yields
$N=4R(LM)^4/32\pi^2+{\rm O}(e^{-bLM})$.
But $N$ is an integer, so the error in this approximation can be
eliminated by taking
\begin{equation}\label{MN}
M_N^4= \frac{32\pi^2N}{4RL^4}.
\end{equation}
This value of $M_N$ is the one needed to cancel the quadratic
divergence.
If one chooses $N=N_{\rm lat}$,
then $M_N^2=4\sqrt{2}\pi/a^2\approx(4.2/a)^2$.

The combination
\begin{equation}\label{main result}
\Gamma_N=\tilde{\Gamma}_N+S_{\rm ct}
\end{equation}
is thus gauge invariant under infinitesimal gauge transformations.
By eq.~(\ref{composition}) this is enough to show that $\Gamma_N(U)$ is
invariant under all lattice gauge transformations.
Hence, $\Gamma_N(U)$ satisfies all three conditions.
This is the main result.

Finally, let us integrate over the zero modes.
Unless there are enough factors of the fermion field in the amplitude,
the integral vanishes.
With the minimal number of fields in eq.~(\ref{B violation})
\begin{equation}\label{amplitudes}
\cZ(n_+,n_-)=e^{-\Gamma_N(A)}
\prod_{n=1}^{n_-}\varphi_{-,n}^{j_n}(y_n)^*
\prod_{n=1}^{n_+}\varphi_{+,n}^{i_n}(x_n),
\end{equation}
where $i_n$, $j_n$ and $x_n$, $y_n$ denote discrete indices and
positions of the fields in eq.~(\ref{B violation}).
Functional integration alone would lead to eq.~(\ref{amplitudes}) with
$\tilde{\Gamma}_N$ instead of $\Gamma_N$.
The zero modes present no substantive changes in the computation of the
gauge variation of $\tilde{\Gamma}_N$, so the same counter-terms restore
gauge symmetry.

\subsection{Relation to ref.~\protect\cite{tHo94}}\label{GtH cutoff}
Although ref.~\cite{tHo94} focuses primarily on the vector-like theory
with Pauli-Villars cutoff, it does prove its important convergence
theorem for theories with vector and axial-vector couplings.
This suggests that the chiral coupling is also intended as an
application.
With the cutoff proposed there, the analysis of cutoff effects leads to
somewhat different conclusions.
Appendix~\ref{PV} recasts 't~Hooft's cutoff in a way that makes
Appendix~\ref{convergence} directly applicable.
One finds that the quadratic divergence drops out, but the universal
term $\alpha_{4R}^a$ still spoils gauge invariance of the real part.
Moreover, the coefficients of the higher-dimension terms,
$\alpha_{6R}^a$ and so forth, no longer vanish, though they are
suppressed by powers of $M^2$.
To eliminate these violations of gauge symmetry, one must take $M$ to
infinity, on each gauge field separately.
This is cumbersome, and perhaps logically inconsistent, because to
obtain a renormalized theory, the bare gauge coupling must be adjusted
to keep physical masses---properties of the ensemble average---cutoff
independent.

\subsection{How to compute baryon violation}\label{how to}
To summarize the results of this section it is worth sketching how to
compute correlation functions.
Because the integrals $\cZ$ sometimes vanish trivially, let us
denote the sector of lattice gauge fields with $n_\pm$ zero modes
$\cU_{(n_+,n_-)}$.
Nonzero integrals with then be over one or so sectors.

Consider first fermion-conserving observables.
One would like to compute a ratio of the form
\begin{equation}\label{observable}
\langle O \rangle=
\frac{\displaystyle\int\cD U_{(0,0)}\; O e^{-\Gamma_N-S_{\rm g}}}
{\displaystyle\int\cD U_{(0,0)}\;e^{-\Gamma_N-S_{\rm g}}},
\end{equation}
where $S_{\rm g}$ is the lattice-gauge-field action.
Here the gauge-invariant observable $O$ is constructed from the gauge
field and fermion propagators.
For eq.~(\ref{observable}) one requires an ensemble of fields in
$\cU_{(0,0)}$, distributed with weight
\begin{equation}\label{weight}
W=e^{-\Gamma_N(U)-S_{\rm g}(U)}.
\end{equation}
For amplitudes of this type, all other sectors carry weight 0, because
of the zero modes, so in the Monte Carlo they are simply omitted.

Consider next a fermion-violating amplitude.
For simplicity, suppose that $O$ does not contain fermion species that
are being created or annihilated.
Now one would like to compute a ratio of the form
\begin{equation}\label{violation}
\langle O\psi^i(x)\psi^j(y)\rangle=
\frac{\displaystyle\int\cD U_{(2,0)}\;
O\varphi_+^i(x)\varphi_+^j(y)e^{-\Gamma_N(U)-S_{\rm g}}}
{\displaystyle\int\cD U_{(0,0)}\;e^{-\Gamma_N(U)-S_{\rm g}}}.
\end{equation}
The numerator and denominator are averages over different sectors,%
\footnote{The two-zero-mode sector $\cU_{(2,0)}$ is typically the sector
with topological charge $Q=1$, which would have one zero mode for each
species.}
and recall that in sectors with zero modes $e^{-\tilde{\Gamma}_N}$ is
defined to be the product of the first $N$ nonzero eigenvalues, ordered
by $\real\lambda_n^2$.
One can re-write eq.~(\ref{violation}) as
\begin{equation}
\langle O\psi^i(x)\psi^j(y)\rangle=
\frac{\displaystyle\int\cD U_{(2,0)}\;e^{-\Gamma_N(U)-S_{\rm g}}}
{\displaystyle\int\cD U_{(0,0)}\;e^{-\Gamma_N(U)-S_{\rm g}}}
\;\langle O\varphi_+^i(x)\varphi_+^j(y)\rangle_{(2,0)},
\end{equation}
where the average $\langle\bullet\rangle_{(2,0)}$ is over $\cU_{(2,0)}$.
In addition to the weight $W$, one must compute the zero-mode
eigenfunctions $\varphi_+(x)$ for created and annihilated species.

The other factor in the fermion-nonconserving amplitude is the ratio of
two partition functions.
To compute it accurately some special numerical techniques
\cite{Bha87,Kar88} are available, which keep track of the system's
preference for one sector or the other.
Various versions of these ``histogram methods'' \cite{Goc88} may also
prove useful in obtaining the nontrivial phase of $W$,
inherent to a complex representation.

\subsection{Global anomalies}
Some theories, the simplest of which is SU(2) with one Weyl doublet,
are afflicted by a global anomaly \cite{Wit82}.
The representations in question are real, and therefore
$e^{-\Gamma_N(U)}$, as defined by eqs.~(\ref{Gamma tilde})
and~(\ref{main result}), is real and positive.
Thus $\Gamma_N(U)$ is~real.

Let us focus on SU(2).
Because $\pi_4({\rm SU(2)})=\zahlen_2$, there are nontrivial gauge
transformations, for which the proof of gauge invariance of
$\Gamma_N(U)$ breaks down.%
\footnote{Although lattice gauge transformations can be built up slowly,
site-by-site, and eq.~(\ref{composition}) shows that the interpolations
inherit this property, the two classes can be separated by lattice
gauge transformations for which the interpolation is ill-defined.}
Let $w$ be in the nontrivial class.
The variation $\Gamma_N(^wU)-\Gamma_N(U)$ is real, but does it vanish?
Following ref.~\cite{Kli91} one can compute the difference by embedding
SU(2) into SU(3) and taking a trajectory from $g=1$ to $g=w$ in SU(3).
$\real\Gamma_N$ does not change for any infinitesimal step, and since
$\pi_4({\rm SU(3)})=0$, the trajectory can be constructed from
infinitesimal steps.
Thus $\Gamma_N(^wU)-\Gamma_N(U)=0$ for the embedded field, and hence
likewise for the SU(2) fields themselves.

On the other hand, Witten argued that the two configurations $U$ and
$^wU$ should have Boltzmann weights equal in magnitude but
{\em opposite\/} in sign \cite{Wit82}.
Indeed the gauge variation of $\imag\Gamma_N$ integrated along the SU(3)
trajectory supports his conclusion \cite{Kli91}.
But, given a lattice gauge field $U'$, the algorithm for
$e^{-\Gamma_N(U')}$ cannot determine whether $U'=U$ or $U'={}^wU$.
And thanks to the pains taken ensure gauge invariance, the computed
weight is the same in either case.
This is not a serious drawback, however.
If the global anomaly applies, one can replace numerator and
denominator of eq.~(\ref{observable}) by Witten's original result,
$0/0$, eliminating all computation.

\section{Conclusions}\label{conclusions}
The seeming incompatibility of lattice fermions and chiral symmetry has
inspired the recurring idea \cite{Flu82,Goe92,tHo94} of treating the
fermions in the continuum, even if the underlying gauge field is on the
lattice.
Smit calls the idea desperate \cite{Smi87}.
How desperate are the main results presented here?
The colorful terminology refers to the tacit assumption that a
continuum requires an infinite number of degrees of freedom (per
unit volume).
Then the arithmetic needed to evaluate functional integrals is infinite:
we are desperate because a computer cannot do the job.
But with a finite-mode cutoff there are $N$ eigenvalues;
the computation is finite.
Alas, even in the vector-like theory the effort needed to obtain
higher eigenvalues will be high (cf.~sect.~\ref{MC}).
The construction for chiral fermions is yet more computationally
intensive, because, first, the counter-terms must now be computed
accurately, and, second, the phase, which stems not from the
regulator but from the chiral coupling itself, requires extra care.

There is a potential shortcoming to the finite-mode regulator.
One would like to verify that the spectrum remains chiral, in
perturbation theory and beyond.
Appendix~\ref{cutoff} demonstrates perturbative universality for fermion
loops.
(Ref.~\cite{And84} and Appendix~\ref{convergence} do the same for
anomalous diagrams only.)
But there is (as yet) no comparable proof for diagrams with external
fermion lines, which are needed to examine the fermion spectrum.
The resolution is not straightforward, because it depends on details of
the gauge-boson propagator, i.e.\ on the lattice gauge action.
And even if the perturbative test is a success, one should be cautious
until the nonperturbative spectrum has been checked.

The counter-terms required in the chiral gauge theory may be unsettling
at first sight.
But their necessity arises from the unassailable observation
\cite{Kar81} that a {\em regulated\/} functional integral either
respects a symmetry or it does not.
To obtain the anomaly (and hence the physically powerful requirement of
anomaly cancellation) the functional integral $e^{-\tilde{\Gamma}_N(U)}$
cannot be gauge symmetric.
Consequently, the final result for the chiral Boltzmann factor
\begin{equation}
e^{-\Gamma_N(U)}= e^{-\tilde{\Gamma}_N(U)} e^{-S_{\rm ct}(U)}
\end{equation}
is the product of a functional integral and a symmetry-restoring
factor.

A related peculiarity is the fate of Witten's global anomaly
\cite{Wit82}.
Again, an Ansatz for the effective action is either gauge invariant or
not.
The effective action $\Gamma_N(U)$ {\em is\/} gauge invariant,
even when the gauge transformation is in the nontrivial class.
Without asserting that this paper's formulation succeeds at defining the
globally anomalous theories, one might suggest that it could shed light
on the dynamical puzzles that originally motivated ref.~\cite{Wit82}.
An optimistic possibility is that the dynamics of $\Gamma_N(U)$ fail in
globally anomalous theories, but not otherwise.

A serious complication is that the formulation is in Euclidean
field theory.
At the nonperturbative level the Wick rotation does no good, and instead
one constructs the Minkowski theory via the imaginary-time evolution
operator on the Hilbert space of states.
This procedure defines a Hamiltonian that can then be used to propagate
the states in real time.
The eigenfunctions used to define the functional integral are
fundamentally four-dimensional, so the constructive approach
\cite{WiK74} does not seem helpful.
Perhaps the axiomatic approach \cite{OSS73} will prove more promising.

Finally, let us compare the present chiral construction with the overlap
formalism \cite{Nar93}.
Both define an effective action with a gauge-invariant real part.
Both generate an imaginary part in a complex fermion representation, but
not in a real representation.
With the interpolation and $\hat{D}$ the gauge variation of the
imaginary part is the consistent anomaly and with the finite-mode
regulator (or in the limit $M\to\infty$ \cite{tHo94}) nothing else.
With the overlap the gauge variation of the imaginary part contains the
anomaly and higher-order gauge breaking terms, analogous to
$\alpha_{6I}^a$ (notation of Appendix~\ref{convergence}), as well.
Ref.~\cite{Nar93} argues they should be tolerably small, and tests in
two dimensions \cite{Nar94} indicate that this conclusion may be
correct.
Both methods provide fermion nonconserving amplitudes: here the
gauge-field topology \cite{Lue82} drives fermion nonconservation,
and in ref.~\cite{Nar93} the fermion nonconserving amplitudes define
the gauge-field topology.
(For smooth lattice gauge fields the two topologies coincide.)
Whereas Appendix~\ref{cutoff} includes an explicit verification of
fermion-loop coupling-constant renormalization, it does not seem that an
explicit calculation starting from the lattice overlap is available yet
\cite{Ran95}.
On the other hand, ref.~\cite{Nar93} includes several numerical cross
checks that have not been done here.
An important, dynamical test is whether the fermion spectrum remains
chiral after integrating over gauge fields.%
\footnote{This is the physical crux of recent skepticism \cite{Gol95}
of the overlap formalism.}
Neither construction has been subjected to this test yet, because it
requires a full-fledged Monte Carlo calculation.

\section*{Acknowledgements}
Sects.~\ref{vector-like}--\ref{metric} are based on an unpublished draft
(1988).
That work was aided by discussions with M. G\"ockeler, G. Schierholz,
and U.-J. Wiese.
I also thank H. Joos for discussions on this topic then and since.
While formulating sect.~\ref{chiral}, I had helpful conversations with
W.A. Bardeen, M.F.L. Golterman, and T. Onogi.

Fermilab is operated by Universities Research Association, Inc.,
under contract DE-AC02-76CH03000 with the U.S. Department of Energy.

\vskip 13pt

While this paper was being written, a preprint appeared by S.D.H. Hsu
(Yale University report YCTP-P5-95, {\tt hep-th/9503058}), expanding on
the idea of G\"ock\-eler and Schierholz \cite{Goe92} to exploit the
$\eta$-invariant formulated in ref.~\cite{Alv86}.

\appendix

\section{Cutoff effects in the anomaly}\label{convergence}
This appendix fills in the steps from eq.~(\ref{regulated anomaly}) to
eq.~(\ref{FdualF}).
The modifications needed to obtain eq.~(\ref{consistent anomaly}) are
provided in sect.~\ref{chiral convergence}.
The analysis is standard, but it is provided to demonstrate the form of
higher-dimension terms suppressed by powers of $1/M_N^{2n}$.
The sharp limit $\varepsilon_N\to0$ turns out to be special, because it
eliminates these terms {\em before\/} taking $M_N\to\infty$.

First some preliminaries on Fourier transforms.
For definiteness the space-time is a box with sides $L$, and volume
$V=L^d$.
The Fourier transform
\begin{equation}\label{Fourier transform}
\tilde{\varphi}(k)=\int d^4x\,e^{-ik\cdot x}\varphi(x).
\end{equation}
Allowed values of $k$ depend on the boundary conditions.
This Appendix presents details for a wide class of almost periodic
boundary conditions,%
\footnote{Other boundary conditions such as Dirichlet, von Neumann, or
fixed lead to the same final conclusions.}
such that $k_\mu=2\pi(\nu_\mu+\eta_\mu)/L$, $\nu_\mu\in\zahlen$.
Strictly periodic directions have $\eta_\mu=0$; anti-periodic directions
have $\eta_\mu=1/2$; $C$-periodic directions have a certain combination
of the foregoing and $\eta_\mu=1/4$.
The inverse transform (assuming convergence) is
\begin{equation}\label{inverse Fourier transform}
\varphi(x)=V^{-1}\sum_k \tilde{\varphi}(k)e^{ik\cdot x}.
\end{equation}
The application needed here is the Fourier transform of
eq.~(\ref{complete})
\begin{equation}\label{k complete}
\sum_n \tilde{\varphi}_n^i(p) \tilde{\varphi}_n^{j*}(q) =
V\delta_{pq}\delta^{ij}.
\end{equation}

Consider any function $f$ obeying $f(0)=1$ and
$f(\infty)=f'(\infty)=f''(\infty)=\cdots=0$ \cite{Fuj79}, and let
\begin{equation}\label{Af-vector}
\cA^a_f(x) = (\gamma_5T^a)_{ji} \sum_n\left[ \varphi^\dagger_n(x)
f(-\slsh{D}^2/M^2) \varphi_n(x)\right]_{ij},
\end{equation}
with the summation convention over the spin-flavor-color multi-indices
$i,j$.
In sect.~(\ref{regulator}) the Fermi function $f_{\varepsilon_N}$
appears, with mass $M_N$ and the limit $\varepsilon_N\to0$.
For comparison with refs.~\cite{tHo94,Fuj79}, however, it is convenient
to keep $f$ arbitrary.

Fourier transforming the eigenfunctions and using eq.~(\ref{k complete})
one obtains
\begin{equation}
\cA^a_f(x) = V^{-1}(\gamma_5T^a)_{ji}\sum_k\left[e^{-ik\cdot x}
f(-\slsh{D}^2/M^2) e^{ik\cdot x}\right]_{ij}.
\end{equation}
  From eq.~(\ref{detC}) one recalls that here
$\slsh{D}e^{ik\cdot x}=e^{ik\cdot x}(i\slsh{k}+\slsh{D})$.
Hence,
\begin{equation}
\cA^a_f(x) = V^{-1}(\gamma_5T^a)_{ji} \sum_k\left[
f\left((k^2-\slsh{D}^2+2ik\cdot D)/M^2\right)\right]_{ij},
\end{equation}
Expanding in $D/M$
\begin{equation}
\cA^a_f(x) = V^{-1}(\gamma_5T^a)_{ji} \sum_k\sum_n
\frac{(-1)^n}{n!}\left(\slsh{D}^2/M^2+2ik\cdot D/M\right)^n_{ij}
f^{(n)}(k^2),
\end{equation}
where now $k_\mu=2\pi(\nu_\mu+\eta_\mu)/(LM)$, and $f^{(n)}=d^nf/dx^n$.

Because the functions under consideration are smooth and vanish rapidly
at infinity, the sums can be approximated by integrals
\begin{equation}\label{integrals}
\frac{1}{(LM)^d} \sum_k (1, k_\mu k_\nu, \ldots) f(k^2) =
\int\frac{d^d k}{(2\pi)^d} (1, k_\mu k_\nu, \ldots) f(k^2) +
{\rm O}(e^{-bLM})
\end{equation}
with little error.
Odd powers vanish.
Note that the finite-size effects also vanish when the ultraviolet
regulator is removed.

Reorganizing the terms according to the power of $M$
\begin{equation}\label{general result}
\begin{array}{r@{}l}
\cA^a_f(x) = & \dfrac{1}{(-4\pi)^{d/2}} \left[
  M^d     f^{( -d/2)}(0)\,\alpha_0^a(x)
+ M^{d-2} f^{(1-d/2)}(0)\,\alpha_2^a(x)
\right. \\[1.0em] + & \left.
  M^{d-4} f^{(2-d/2)}(0)\,\alpha_4^a(x)
+ M^{d-6} f^{(3-d/2)}(0)\,\alpha_6^a(x) +\cdots\right],
\end{array}
\end{equation}
where $\alpha_0^a(x)=\tr\gamma_5 T^a=0$,
\begin{equation}
\alpha_2^a = - \tr[\gamma_5T^a(\slsh{D}^2-D^2)],
\end{equation}
\begin{equation}
\alpha_4^a = \half \tr\left\{\gamma_5T^a
\left((\slsh{D}^2-D^2)^2 + \third[D_\mu,[D_\mu,\slsh{D}^2-D^2]]
+\sixth[D_\mu,D_\nu][D_\mu,D_\nu] \right)\right\},
\end{equation}
and $\alpha_6^a$ contains terms with 6 $D$'s, combined to produce a
function (rather than a differential operator).
The result depends on the cutoff function via the coefficients
$f^{(n)}(0)$, defined by $f^{(0)}(x)\equiv~f(x)$,
\begin{equation}
f^{(n-1)}(x)=-\int^\infty_x dx'f^{(n)}(x'), \quad
f^{(n+1)}(x)=\frac{df^{(n)}}{dx}.
\end{equation}
The $M$ independent term has the universal coefficient $f(0)=1$;
the other coefficients differ for different cutoffs.

With the Fermi function $f_\varepsilon^{(-n)}(0)=(-1)^n/n!$, $n>0$,
plus terms of order $\varepsilon^{n}e^{-1/\varepsilon}$.
Hence, the $\alpha_{2j}^a$, $2j<d$, are power-law divergences, unless
the traces vanish.
On the other hand, for $n>0$ then
$f_\varepsilon^{(n)}(0)\sim\varepsilon^{-n}e^{-1/\varepsilon}\to0$
as $\varepsilon\to0$.
Thus, the sharp cutoff has no ``scaling violations.''

For the vector-like theory
$\slsh{D}^2-D^2=\quart[\gamma_\mu,\gamma_\nu]F_{\mu\nu}$.
If $d=2$, the only surviving term $\alpha_2^a$ yields the well-known
result.
If $d=4$, the Dirac trace makes $\alpha_2^a$ vanish as well as
everything in $\alpha_4^a$ except the term $(\slsh{D}^2-D^2)^2$,
yielding the familiar axial anomaly, eq.~(\ref{FdualF}).

\subsection{Modifications for the chiral gauge
theory}\label{chiral convergence}
In the chiral gauge theory one wants
\begin{equation}\label{Af-chiral}
\cA^a_f(x) = (\gamma_5t^a)_{ji} \sum_n\left[ \chi^\dagger_n(x)
f(-\hat{D}^2/M^2) \eta_n(x)\right]_{ij}.
\end{equation}
The left and right eigenfunctions of $i\hat{D}$ are not complete, but
\begin{equation}
\sum_n \eta_n^i(x) \chi_n^{j*}(y) = \delta(x-y)\delta^{ij}
- {\sf I\kern-0.15em P}_0,
\end{equation}
where ${\sf I\kern-0.15em P}_0$ projects onto zero modes.
This nuisance is easiest to handle with anti-periodic boundary
conditions.
If the function $f$ drops to zero below the lowest momentum mode,
then ${\sf I\kern-0.15em P}_0f=0$.
Thus, the projector can be dropped after Fourier transforming.
The hole in $f$ cannot affect the momentum sums, so the correct
approximation in eq.~(\ref{integrals}) uses a function without the hole
on the right-hand side.

The algebraic manipulations still hold, but one must replace $T^a$ by
$t^a$, $\slsh{D}$ by $\hat{D}$ {\em and\/} the symbol $D_\mu$
by $\bar{D}_\mu:=\half\{\gamma_\mu,\hat{D}\}$.
  From eq.~(\ref{D hat})
\begin{equation}
\bar{D}_\mu=
\partial_\mu + \half A_\mu - \ihalf \sigma_{\mu\nu}\gamma_5 A_\nu.
\end{equation}
Then
\begin{equation}
\hat{D}^2-\bar{D}^2=
\half \gamma_5 \partial\vdot A
- \ihalf \sigma_{\mu\nu} \partial_\mu A_\nu
+\quart(d-2)(A^2- i\sigma_{\mu\nu} A_\mu A_\nu).
\end{equation}
Consequently,
\begin{equation}
\alpha_2^a=\left\{
\begin{array}{ll}
\varepsilon_{\mu\nu}\tr(t^a \partial_\mu A_\nu)
- \tr(t^a \partial\vdot A) & (d=2) \\
-2\tr(t^a \partial\vdot A) & (d=4),
\end{array}
\right.
\end{equation}
where the trace is now only over gauge indices.
In $d=2$ this is the universal anomaly, plus a term that can be
compensated by a local counter-term;
in $d=4$ it only the latter survives, and it is a power-law divergence.
The function in $\alpha_4^a$ is too lengthy to present explicitly,
but after the trace $\alpha_4^a=\alpha_{4R}^a+\alpha_{4I}^a$, where
\begin{equation}
\alpha_{4I}^a=\tfrac{2}{3}\varepsilon_{\mu\nu\rho\sigma}
\tr\left[t^a\partial_\mu(A_\nu\partial_\rho A_\sigma
+ \half A_\nu A_\rho A_\sigma)\right]\quad(d=4)
\end{equation}
is the well-known consistent anomaly \cite{Bar69,Wes71,Gro72}, and
\begin{equation}\label{real part}
\begin{array}{r@{}l}
\alpha_{4R}^a=\third\tr&\left[t^a\left(
  \partial^2\partial\vdot A + [\partial^2 A_\mu, A_\mu]
- 2 [\partial_\mu \partial\vdot A, A_\mu] + \{\partial\vdot A, A^2\}
\rule{0.0em}{0.9em}\right.\right. \\[1.0em]
  & \left.\left.\rule{0.0em}{0.9em}
- A_\mu(\partial\vdot A)A_\mu - [\partial_\mu A_\nu, [A_\mu, A_\nu]]
+ A_\mu(\partial_\mu A_\nu + \partial_\nu A_\mu)A_\nu
\right)\right]
\end{array}
\end{equation}
is the almost as well-known quantity that, like $\alpha_2^a|_{d=4}$,
can be compensated by local counter-terms \cite{Bar69,Bal82,And84}.

\section{Cutoff effects in the effective action}\label{cutoff}
The analysis of the previous section can be applied directly to the
effective action.
It shows that the finite-mode cutoff is in the same (perturbative)
universality class as Pauli-Villars regulators.
As in Appendix~\ref{convergence}, the analysis is performed for an
arbitrary smooth function, but once again the sharp limit is special,
because it has no power corrections.

For a good infrared behavior, this section considers only anti-periodic
boundary conditions.

One can write the regulated effective action as
\begin{equation}\label{Gamma_f}
\Gamma_N = -\half \sum_n^\infty
\log\left(\frac{\lambda_n^2+m^2}{M_N^2}\right)
f_{\varepsilon_N}\left(\frac{\lambda_n^2}{M_N^2}\right)=
\int d^4x\,\cL_N,
\end{equation}
where the effective Lagrangian
\begin{equation}\label{lagrangian}
\cL_N=-\half\sum_n\varphi^\dagger_n(x)
L_{f_{\varepsilon_N}}(-\slsh{D}^2/M_N^2)\varphi_n(x),
\end{equation}
and, for any function $f$ obeying $f(0)=1$ and
$f(\infty)=f'(\infty)=f''(\infty)=\cdots=0$,
\begin{equation}\label{L_f}
L_f(x) = \log(x+\mu^2)f(x),
\end{equation}
and $\mu=m/M$.
The notation applies to the vector-like theory, but the substitutions
needed for the chiral gauge theory are obvious.
The effective Lagrangian for arbitrary $f$ will be denoted $\cL_f$.

Fourier transforming the eigenfunctions, using eq.~(\ref{k complete}),
and using
$\slsh{D}e^{ik\cdot x}=e^{ik\cdot x}(i\slsh{k}+\slsh{D})$ one obtains
\begin{equation}
\cL_f(x) = -\half\delta_{ji}V^{-1} \sum_k\left[
L_f\left((k^2-\slsh{D}^2+2ik\cdot D)/M^2\right)\right]_{ij},
\end{equation}
where $k_\mu=\pi(2\nu_\mu+1)/L$.
Expanding in $D/M$
\begin{equation}
\cL_f(x) = -\half\delta_{ji}V^{-1} \sum_k\sum_n
\frac{(-1)^n}{n!}\left(\slsh{D}^2/M^2+2ik\cdot D/M\right)^n_{ij}
L_f^{(n)}(k^2),
\end{equation}
where $k_\mu=\pi(2\nu_\mu+1)/(LM)$.
The derivatives take the form
\begin{equation}\label{Lfn}
L_f^{(n)}(x)= \sum_{l=0}^{n-1} \frac{n!}{l!} \frac{(-1)^{n-1-l}}{(n-l)}
\frac{f^{(l)}(x)}{(x+\mu^2)^{n-l}} + \log(x+\mu^2)f^{(n)}(x).
\end{equation}
The $l=0$ term can be called universal, because after summing over
$k$ the universal $f(0)=1$ remains.

In the vector-like theory, the fermion mass (presumed nonzero) regulates
the infrared and one can use eq.~(\ref{integrals}).
In the chiral gauge theory (or any massless case), infrared
singularities make the integrals poor estimates of the sums.
For the present purposes, however, the function $f$ can also be used as
an infrared regulator.
One simply chooses $f$ to drop to zero for momentum less than the
smallest allowed by the anti-periodic boundary condition, say for
$k^2<\delta^2=(\pi/2LM)^2$.

The delicate infrared behavior has consequences when reorganizing the
effective Lagrangian according to the dimension of the interactions.
One finds
\begin{equation}\label{lagrangian expansion}
\begin{array}{r@{}l}
\cL_f = & \dfrac{-1}{2(-4\pi)^{d/2}} \left[
  M^d     L_f^{( -d/2)}(\delta^2)\,l_0(x)
+ M^{d-2} L_f^{(1-d/2)}(\delta^2)\,l_2(x)
\right. \\[1.0em] + & \left.
  M^{d-4} L_f^{(2-d/2)}(\delta^2)\,l_4(x)
+ M^{d-6} L_f^{(3-d/2)}(\delta^2)\,l_6(x) +\cdots\right]+\Delta\cL,
\end{array}
\end{equation}
where $l_0(x)=\tr1=4R$ yields a constant of no dynamical significance,
\begin{equation}
l_2 = - \tr(\slsh{D}^2-D^2),
\end{equation}
\begin{equation}
l_4 = \half \tr\left\{
(\slsh{D}^2-D^2)^2 + \third[D_\mu,[D_\mu,\slsh{D}^2-D^2]]
+\sixth[D_\mu,D_\nu][D_\mu,D_\nu]\right\},
\end{equation}
and $l_6$ contains same combination of 6 $D$'s as $\alpha_6^a$,
which is a function rather than a differential operator.
The interactions in $\Delta\cL$ are proportional to
$[\delta^2/(\delta^2+\mu^2)]^l$;
hence they survive only when $\mu^2\ll\delta^2$.

The effective Lagrangian depends on the cutoff via the coefficients
$L_f^{(n)}(\delta^2)$.
If $\delta^2\ll\mu^2$ (i.e.~$mL\gg1$) these become $L_f^{(n)}(0)$.
For example, with the Fermi function
\begin{equation}
L_{f_\varepsilon}^{(-1)}(0)=1, \quad L_{f_\varepsilon}^{(-2)}(0)=-1/4,
\end{equation}
plus terms of order $\varepsilon^{|n|}e^{-1/\varepsilon}$.
The higher-dimension functions have coefficients
$L_f^{(n)}(\delta^2)/M^{2n}$, $n>0$.
  From eq.~(\ref{Lfn}) one sees that they consist of a universal,
$f$- and $M$-independent term, plus terms non-universal proportional
to $f^{(m)}(0)/M^{2m}$.
The remaining dimensions are balanced by infrared scales $m$ or $1/L$.
Again, for the Fermi function the non-universal terms are suppressed by
$e^{-1/\varepsilon}$ and drop out in the sharp limit.
Since interactions of all dimension appear in
eq.~(\ref{lagrangian expansion}), it must be interpreted as a
perturbative series.

In the vector-like theory, regulated as in sect.~\ref{regulator},
$l_2$ vanishes.
Indeed, the gauge invariance of the {\em regulated\/} theory forbids any
dimension 2 terms.
In $l_4$ the Dirac trace eliminates the nested commutator leaving,
for $d=4$,
\begin{equation}
\cL_{f,4}=\frac{\log(\mu^2,\delta^2)}{48\pi^2} {\tr}_\rho F^2,
\end{equation}
which renormalizes the gauge coupling.

In the chiral gauge theory of sect.~\ref{chiral} this analysis applies
to $\tilde{\Gamma}_N$.
The dimension-two term $l_2$ does not vanish, because now the
regulator breaks the gauge symmetry.
In four dimensions one has
\begin{equation}
\cL_{f,2}=\frac{M^2}{16\pi^2}L_f^{(-1)}(\delta^2){\tr}_\rho A^2.
\end{equation}
To restore gauge symmetry, one must add the counter-term $S_2$ in
eq.~(\ref{counter-terms}).
On the other hand, $l_4$ induces coupling constant renormalization.
(The gauge non-invariant pieces cancel.)
With some patience one can accumulate the nonvanishing $\gamma$-matrix
traces in $l_4$ to obtain
\begin{equation}
\cL_{f,4}=\frac{\log\delta^2}{96\pi^2} {\tr}_\rho F^2,
\end{equation}
up to a total derivative.
Notice that, as expected, the renormalization term is half that of the
vector-like theory.
The dimension-four interactions in $\Delta\cL$, present because
$\mu^2=0$, include gauge-breaking terms; they are cancelled by $S_4$.

\section{Applying
Appendices~\protect\ref{convergence} and~\protect\ref{cutoff}
to ref.~\protect\cite{tHo94}}\label{PV}
Ref.~\cite{tHo94} defines the fermion functional integral
with a time-honored \cite{tHo71} Pauli-Villars regulator:
\begin{equation}
e^{-\Gamma_{\rm PV}(A)}=\prod_i\left(\det(\slsh{D}+M_i)\right)^{e_i}
\end{equation}
in the vector-like case, and similarly $e^{-\tilde{\Gamma}}$ with
$\hat{D}$ in the chiral case.
The determinants are products over the infinitely numerous eigenvalues.
The masses%
\footnote{$M_0=m$ for vector-like gauge theories;
$M_0=0$ for chiral gauge theories.}
$M_i$ and signatures $e_i$ satisfy
\begin{equation}\label{powers}
\sum_{i=0}^\infty e_i =
\sum_{i=0}^\infty e_i (M_i/M)^n = 0,\quad n=1,2,3,\ldots
\end{equation}
\begin{equation}\label{logs}
\sum_{i=1}^\infty e_i (M_i/M)^n \log(M_i/M) = 0,\quad n=0,1,2,3,\ldots
\end{equation}
For $n=0$, eq.~(\ref{logs}) defines the overall scale $M$.
Eqs.~(\ref{powers}) and (\ref{logs}) require infinite series only if one
requires the identities for all $n$.

The function needed to describe the effective Lagrangian,
cf.\ eqs.~(\ref{Gamma_f})--(\ref{lagrangian expansion}), is
\begin{equation}
L_{\rm PV}(x)=\sum_{i=0}^\infty e_i\log(x + M_i^2/M^2).
\end{equation}
Note that $L_{\rm PV}$ and all its derivatives vanish at infinity,
by virtue of eqs.~(\ref{powers}) and~(\ref{logs}),
so the manipulations of Appendix~\ref{cutoff} still hold.
Eqs.~(\ref{aWTi}) and (\ref{gauge variation}) also hold as before,
but with $\cA_{\rm reg}$ replaced by $\cA_{f_{\rm PV}}$.
Here
\begin{equation}
f_{\rm PV}(x)=xL'_{\rm PV}(x)=
-\sum_{i=1}^\infty \frac{e_iM_i^2}{M^2x + M_i^2}.
\end{equation}
Again $f_{\rm PV}$ and all its derivatives vanish at infinity, so
the manipulations of Appendix~\ref{convergence} hold.

To apply eq.~(\ref{general result}) one notes the universal
normalization $f_{\rm PV}(0)=1$.
Power-law divergences drop out,
but ``scaling violations'' remain ($n>0$):
\begin{equation}
f_{\rm PV}^{(-n)}(0)=0,\quad
f_{\rm PV}^{(n)}(0)=
\frac{(-1)^{n+1}}{n!}\sum_{i=1}^\infty e_i (M/M_i)^{2n} \neq0,
\end{equation}
and analogously for $L^{(n)}_{\rm PV}(0)$.
One might remark that the absence of power-law divergences relies on
integrating $x$ to infinity.
They re-appear if one truncates $\Gamma_{\rm PV}$ when, say,
$L_{\rm PV}(x)$ becomes small.
\pagebreak[1]


\begin{thebibliography}{99}
\bibitem{Kar81}
L.H. Karsten and J. Smit, Nucl. Phys. {\bf B183} (1981) 103.
\bibitem{Nie81}
H.B. Nielsen and H. Ninomiya, Nucl. Phys. {\bf B185} (1981) 20,
(E) {\bf B195} (1982) 541; {\bf B193} (1981) 173;\\
D. Friedan, Commun. Math. Phys. {\bf 85} (1982) 481.
\bibitem{Wil77}
K.G. Wilson, in {\em New Phenomena in Subnuclear Physics\/},
edited by A. Zichichi (Plenum, New York, 1977).
\bibitem{Sus77}
L. Susskind, Phys. Rev. {\bf D16} (1977) 3031;\\
T. Banks, J.B. Kogut, and L. Susskind, Phys. Rev. {\bf D13} (1976) 1043.
\bibitem{Pet93}
D.N. Petcher, Nucl. Phys. B (Proc. Suppl.) {\bf 30} (1993) 50.
%
\bibitem{Nar93}
R. Narayanan and H. Neuberger, % Phys. Lett. {\bf B302} (1993) 62;
Phys. Rev. Lett. {\bf 71} (1993) 3251;
Nucl. Phys. {\bf B412} (1994) 574;
Institute for Advanced Study report IASSNS-HEP-94/99
({\tt hep-th/9411108}).
\bibitem{Kap92}
D.B. Kaplan, Phys. Lett. {\bf B288} (1992) 342.
\bibitem{Fro94}
S.A. Frolov and A.A. Slavnov, Phys. Lett. {\bf B309} (1993) 344;
Nucl. Phys. {\bf B411} (1994) 647;\\
R. Narayanan and H. Neuberger,  Phys. Lett. {\bf B302} (1993) 62.
%
\bibitem{Flu82}
R. Flume and D. Wyler, Phys. Lett. {\bf 108B} (1982) 317.
\bibitem{tHo94}
G. 't~Hooft, Utrecht University report THU-94/18 ({\tt hep-th/9411228}).
%
\bibitem{Ati71}
M. Atiyah and I.M. Singer, Ann. Math. {\bf 93} (1971) 139.
%
\bibitem{Kro87}
A.S. Kronfeld, Nucl. Phys. B (Proc. Suppl.) {\bf 4} (1988) 329.
\bibitem{Smi87}
J. Smit, Nucl. Phys. B (Proc. Suppl.) {\bf 4} (1988) 451.
\bibitem{Goe92}
M. G\"ockeler and G. Schierholz,
Nucl. Phys. B (Proc. Suppl.) {\bf 29} (1992) 114; {\bf 30} (1993) 609.
%
\bibitem{And84}
A. Andrianov and L. Bonora, Nucl. Phys. {\bf B233} (1984) 232, 247.
%
\bibitem{Fuj79}
K. Fujikawa, Phys. Rev. Lett. {\bf 42} (1979) 1195;
Phys. Rev. {\bf D21} (1980) 2848, (E) {\bf D22} (1980) 1499.
%
\bibitem{Lue82}
M. L\"uscher, Comm. Math. Phys. {\bf 85} (1982) 29.
\bibitem{Phi86}
A. Phillips and D. Stone, Comm. Math. Phys. {\bf 103} (1986) 599.
\bibitem{Goe93}
M. G\"ockeler, A.S. Kronfeld, G. Schierholz, and U.-J. Wiese,
Nucl. Phys. {\bf B404} (1993) 839.
\bibitem{Kro88}
A.S. Kronfeld, unpublished (1988).
\bibitem{vBa82}
P. van Baal, Comm. Math. Phys. {\bf 85} (1982) 529.
\bibitem{Goe87}
M. G\"ockeler, A.S. Kronfeld, M.L. Laursen, G. Schierholz,
and U.-J. Wiese, Nucl. Phys. {\bf B292} (1987) 349.
% Phys. Lett. {\bf B209} (1988) 315.
%
\bibitem{Wie93} % fixed-point action for free fermions
U.-J. Wiese, Phys. Lett. {\bf B315} (1993) 417.
%
\bibitem{tHo76}
G. 't~Hooft, Phys. Rev. Lett. {\bf 37} (1976) 8;
Phys. Rev. {\bf D14} (1976) 3432, (E) {\bf D18} (1978) 2199.
%
\bibitem{Alv86}
L. Alvarez-Gaum\'e and S. Della Pietra, in {\em Recent Developments in
Field Theory and Statistical Mechanics}, edited by J. Ambj\o rn, B.J.
Durhuus, and J.L Petersen (Elsevier, Amsterdam, 1985);\\
L. Alvarez-Gaum\'e, S. Della Pietra, and V. Della Pietra,
Phys. Lett. {\bf 166B} (1986) 177;\\
S. Della Pietra, V. Della Pietra, and L. Alvarez-Gaum\'e,
Commun. Math. Phys. {\bf 109} (1987) 691.
\bibitem{Alv84}
L. Alvarez-Gaum\'e and P. Ginsparg, Nucl. Phys. {\bf B243} (1984) 449.
%
\bibitem{Bal82}
A.P. Balachandran, G. Marmo, V.P. Nair, and C.G. Trahern,
Phys. Rev. {\bf D25} (1982) 2713.
\bibitem{Ein84}
M.B. Einhorn and D.R.T. Jones, Phys. Rev. {\bf D29} (1984) 331.
%
\bibitem{Bar69}
W.A. Bardeen, Phys. Rev. {\bf 184} (1969) 1848.
\bibitem{Wes71}
J. Wess and B. Zumino, Phys. Lett. {\bf 37B} (1971) 95.
\bibitem{Gro72}
D.J. Gross and R. Jackiw, Phys. Rev. {\bf D6} (1972) 477.
%
\bibitem{Bor90}
A. Borelli, L. Maiani, G.C. Rossi, R. Sisto, and M. Testa,
Nucl. Phys. {\bf B333} (1990) 335.
\bibitem{Alo90}
J.L. Alonso, Ph. Boucaud, J.L. Cort\'es, and E. Rivas,
Mod. Phys. Lett. {\bf A5} (1990) 275; Phys. Rev. {\bf D44} (1991) 3258.
\bibitem{Bod91}
G.T. Bodwin and E.V. Kov\'acs,
Nucl. Phys. B (Proc. Suppl.) {\bf 20} (1991) 546; {\bf 30} (1993) 617.
%
\bibitem{Bha87}
G. Bhanot, S. Black, P. Carter, and R. Salvador,
Phys. Lett. {\bf B183} (1987) 331;\\
G. Bhanot, K. Bitar, and R. Salvador,
Phys. Lett. {\bf B187} (1987) 381; {\bf B188} (1988) 246;\\
G. Bhanot, A. Gocksch, and P. Rossi, Phys. Lett. {\bf B199} (1988) 101.
\bibitem{Kar88}
M. Karliner, S. Sharpe, and Y.F. Chang,
Nucl. Phys. {\bf B302} (1988) 204.
\bibitem{Goc88}
A. Gocksch, Phys. Lett. {\bf B206} (1988) 290;\\
U.-J. Wiese, Nucl. Phys. {\bf B318} (1989) 153.
%
\bibitem{Wit82}
E. Witten, Phys. Lett. {\bf 117B} (1982) 324.
\bibitem{Kli91}
F. Klinkhamer, Phys. Lett. {\bf B256} (1991) 41.
%
\bibitem{WiK74}
K.G. Wilson and J. Kogut, Phys. Reports {\bf 12C} (1974) 75;\\
M. L\"uscher, Commun. Math. Phys. {\bf 54} (1977) 283.
\bibitem{OSS73}
K. Osterwalder and R. Schrader,
Commun. Math. Phys. {\bf 31} (1973) 83, {\bf 42} (1975) 281;\\
K. Osterwalder and E. Seiler, Ann. Phys. {\bf 110} (1978) 440.
%
\bibitem{Nar94}
R. Narayanan and H. Neuberger, Rutgers University report RU-94-99
({\tt hep- lat/9412104});\\
R. Narayanan, H. Neuberger, and P. Vranas,
Institute for Advanced Study report IASSNS-HEP-95-19
({\tt hep-lat/9503013}).
\bibitem{Ran95}
S. Randjbar-Daemi and J. Strathdee, ICTP (Trieste) report IC-95-6
({\tt hep-lat/ 9501027}).
\bibitem{Gol95}
M.F.L. Golterman and Y. Shamir, Institute for Theoretical Physics report
NSF-ITP-95-60 ({\tt hep-lat/9501035}).
%
\bibitem{tHo71}
G. 't~Hooft, Nucl. Phys. {\bf B33} (1971) 173.
\end{thebibliography}
\end{document}